\documentclass[twocolumn,preprintnumbers,amsmath,amssymb]{revtex4-1}

\usepackage{graphicx}              
\usepackage{dcolumn}               
\usepackage{longtable}             
\usepackage{bm}                    
\usepackage{afterpage}
\usepackage{color}

\def\beq{\begin{equation}}
\def\eeq{\end{equation}}

\begin{document}

\title{Approaching Chemical Accuracy with Quantum Monte Carlo}

\author{F. R. Petruzielo$^{1}$}
\email{frp3@cornell.edu}
\author{Julien Toulouse$^{2}$}
\email{julien.toulouse@upmc.fr}
\author{C. J. Umrigar$^{1}$}
\email{CyrusUmrigar@cornell.edu}
\affiliation{
$^1$Laboratory of Atomic and Solid State Physics, Cornell University, Ithaca, New York 14853, USA\\
$^2$Laboratoire de Chimie Th\'eorique, Universit\'e Pierre et Marie Curie and CNRS, 75005 Paris, France\\
}

\date{\today}
\begin{abstract}
A quantum Monte Carlo study of the atomization energies for the G2 set of molecules is presented.
Basis size dependence of diffusion Monte Carlo atomization energies is studied with a single determinant Slater-Jastrow trial wavefunction formed from Hartree-Fock orbitals.
With the largest basis set, the mean absolute deviation from experimental atomization energies for the G2 set is 3.0 kcal/mol.
Optimizing the orbitals within variational Monte Carlo improves the agreement between diffusion Monte Carlo and experiment, reducing the mean absolute deviation to 2.1 kcal/mol.
Moving beyond a single determinant Slater-Jastrow trial wavefunction, diffusion Monte Carlo with a small complete active space Slater-Jastrow trial wavefunction results in near chemical accuracy.
In this case, the mean absolute deviation from experimental atomization energies is 1.2 kcal/mol.
It is shown from calculations on systems containing phosphorus that the accuracy can be further improved by employing a larger active space.
\end{abstract}
\maketitle
\section{Introduction}
\label{sec:intro}
Quantum Monte Carlo (QMC) \cite{Foulkes2001} is touted by devotees as a ``very accurate'' method.
However, previous QMC studies of the atomization energies of the molecules in the G2 set \cite{Curtiss1991} have not obtained chemical accuracy \cite{Grossman2002,Nemec2010}, defined as 1 kcal/mol.
These studies, which are limited to a single determinant Slater-Jastrow (SJ) trial wavefunction and a fixed set of orbitals obtained via a quantum chemistry calculation,
produce a mean absolute deviation (MAD) from experimental atomization energies of about 3 kcal/mol.

This work aims to improve upon both of those shortcomings.
As a starting point, a single determinant SJ trial wavefunction composed of Hartree-Fock (HF) orbitals is used to compute the diffusion Monte Carlo (DMC) \cite{Umrigar1993} atomization energies for the G2 set.
These calculations, which are performed for double-zeta ($2z$), triple-zeta ($3z$), and quintuple-zeta ($5z$) bases, demonstrate the convergence of the DMC atomization energies with respect to basis size.
The MAD from experiment for the $5z$ basis is 3.0 kcal/mol, in agreement with previous QMC studies \cite{Grossman2002,Nemec2010}.

Next, the restriction to a fixed set of quantum chemistry orbitals is relaxed.
The orbitals for each system and basis are optimized in variational Monte Carlo (VMC) via the linear method \cite{Toulouse2007,Toulouse2008,Umrigar2007}.
Employing the single determinant SJ trial wavefunction with optimized orbitals, DMC yields a MAD from experiment of 2.1 kcal/mol for the $5z$ basis.

Finally, the restriction of a single determinant SJ trial wavefunction is relaxed.
With a complete active space (CAS) SJ trial wavefunction formed from just an $s$ and $p$ valence orbital active space, DMC produces atomization energies of near chemical accuracy.
The MAD from experimental atomization energies is 1.2 kcal/mol.
This lends some backing to the claim that QMC is ``very accurate''.
It is found that the MAD can be further reduced by including valence $d$ orbitals in the active space for the heavier systems that are underbound in DMC.

This paper is organized as follows.
In Section \ref{sec:setup}, the computational setup is described.
In Section \ref{sec:results}, results of the computations are described.
Concluding remarks are in Section \ref{sec:conclusion}.
\section{Computational Setup}
\label{sec:setup}
All QMC calculations performed for this work use the Burkatzki-Filippi-Dolg (BFD) pseudopotentials \cite{Burkatzki2007,BFD} in the QMC package CHAMP \cite{CHAMP}. 
The $2z$ and $3z$ basis sets are the recently developed atomic natural orbital Gauss-Slater (ANO-GS) bases \cite{Petruzielo2010, Petruzielo2011}.
For the $5z$ basis, the Gaussian BFD basis set \cite{Burkatzki2007,BFD} is used, omitting the $g$ and $h$ functions.
In the course of this study, it was determined that the hydrogen pseudopotential produced unreliable atomization energies.
A significantly improved pseudopotential for hydrogen was developed by Filippi and Dolg, and is used in this work.
Also, $2z$ and $3z$ ANO-GS basis sets, and a $5z$ Gaussian basis set appropriate for this pseudopotential have been constructed for this work.
The improved pseudopotential and corresponding basis sets are available in the supplementary material \cite{supplementary}.

A combination of experimental and theoretical molecular geometries are used in this study \cite{Feller2008, Oneill2005, CCCBDB, Kalemos2004}.
The zero point energies and experimental atomization energies are from Feller et al. \cite{Feller1999, Feller2008}.
The geometries, zero point energies, and experimental atomization energies for each molecule are available in the supplementary material \cite{supplementary}.

For single determinant SJ trial wavefunctions, the initial orbitals are generated in GAMESS \cite{Schmidt1993} via spin-restricted Hartree-Fock calculations.
The Jastrow parameters, and when applicable, the orbital parameters, are then optimized in VMC via the linear method \cite{Toulouse2007,Toulouse2008,Umrigar2007}.

For CAS SJ trial wavefunctions, the initial orbitals and initial configuration state function (CSF) coefficients are generated in GAMESS via multi-configurational self-consistent field theory (MCSCF) calculations.
The Jastrow, orbital, and CSF parameters are then optimized in VMC via the linear method.
The active space consists of the $1s$ orbital for hydrogen,
the $2s$ and $2p$ orbitals for the first row atoms, and
the $3s$ and $3p$ orbitals for the second row atoms, and the corresponding orbitals for the molecules.

Additionally, not all of the CSFs generated by the MCSCF calculations are included in the QMC calculations.
Instead, a dual criterion for selecting CSFs is employed.
If the magnitude of a CSF coefficient is at least 0.005 or a CSF is a double excitation from the HF CSF, then it is included in the trial wavefunction.
This dual criterion is employed in contrast to the usual single criterion based only on the magnitude of CSF coefficients because the optimal CSF coefficients in QMC can differ greatly from the coefficients generated
via MCSCF.  
Although the magnitude of most CSF coefficients decrease upon optimization in VMC due to the Jastrow factor's effectiveness in describing electronic correlations, 
there are systems for which the magnitude of the coefficients for a few double excitations increase considerably.
This dual selection criterion results in a relatively modest number of CSFs.
The largest number employed is for C$_2$H$_6$ and Si$_2$H$_6$.
These trial wavefunctions consist of 650 CSFs comprising 1700 unique determinants, whereas the MCSCF calculation generates 1.4 million CSFs.

Finally, all DMC calculations are performed with a $0.01$ H$^{-1}$ time step.
The walker populations are large enough for a negligible population control bias and furthermore the small population control bias is
eliminated using the method described in Refs. \cite{NightingaleBloete86,Umrigar1993}.
For all systems except LiH, BeH, CH$2$ ($^3$B$_1$), LiF, C$_2$H$_2$, CN, HCN, HCO, NaCl the locality approximation \cite{Mitas1991} is employed for the nonlocal pseudopotential.
The aforementioned systems suffer from instabilities with the locality approximation, so those computations are performed with the size-consistent version of the T-moves approximation \cite{Casula2010}.
Note that for these systems the atomic energies are also calculated with T-moves so that atomization energies are always calculated in a consistent manner.
All DMC calculations are performed with a sufficient number of Monte Carlo steps such that the statistical error bar on the atomization energy of each system is about ~0.1 kcal/mol.
\section{Results}
\label{sec:results}
The raw data for all calculations presented here are available in the supplementary material \cite{supplementary}.

\begin{figure*}[htp]  
 \begin{center}
   \includegraphics[scale=0.45]{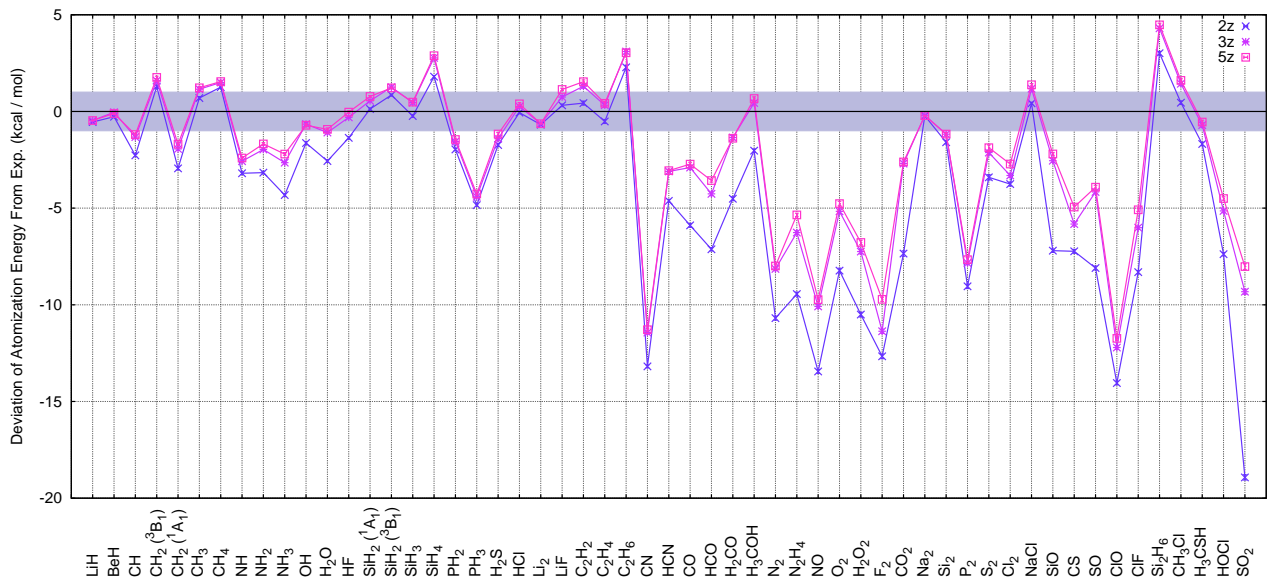}
   \caption{Deviation of the DMC atomization energies from experiment for a single determinant SJ trial wavefunction composed of HF orbitals.
   The MAD from experiment for the $2z$, $3z$, and $5z$ bases are 4.5 kcal/mol, 3.2 kcal/mol, and 3.0 kcal/mol, respectively.}
   \label{fig:j}
 \end{center}
\end{figure*}

The deviation of the DMC atomization energies from experiment for a single determinant SJ trial wavefunction composed of HF orbitals is shown in Figure \ref{fig:j}.
The results for $2z$, $3z$, and $5z$ basis sets demonstrate the convergence of the atomization energies with respect to basis size.
The MAD from experiment for the $2z$, $3z$, and $5z$ bases are 4.5 kcal/mol, 3.2 kcal/mol, and 3.0 kcal/mol, respectively.
The $5z$ result agrees with previous QMC studies \cite{Grossman2002,Nemec2010} which had a MAD from experiment of about 3 kcal/mol.
Note that Nemec et. al performed all-electron DMC calculations with HF orbitals \cite{Nemec2010} whereas Grossman employed the Stevens-Basch-Krauss pseudopotentials \cite{Stevens1984} with MCSCF natural orbitals \cite{Grossman2002}.

\begin{figure*}[htp]  
 \begin{center}
   \includegraphics[scale=0.45]{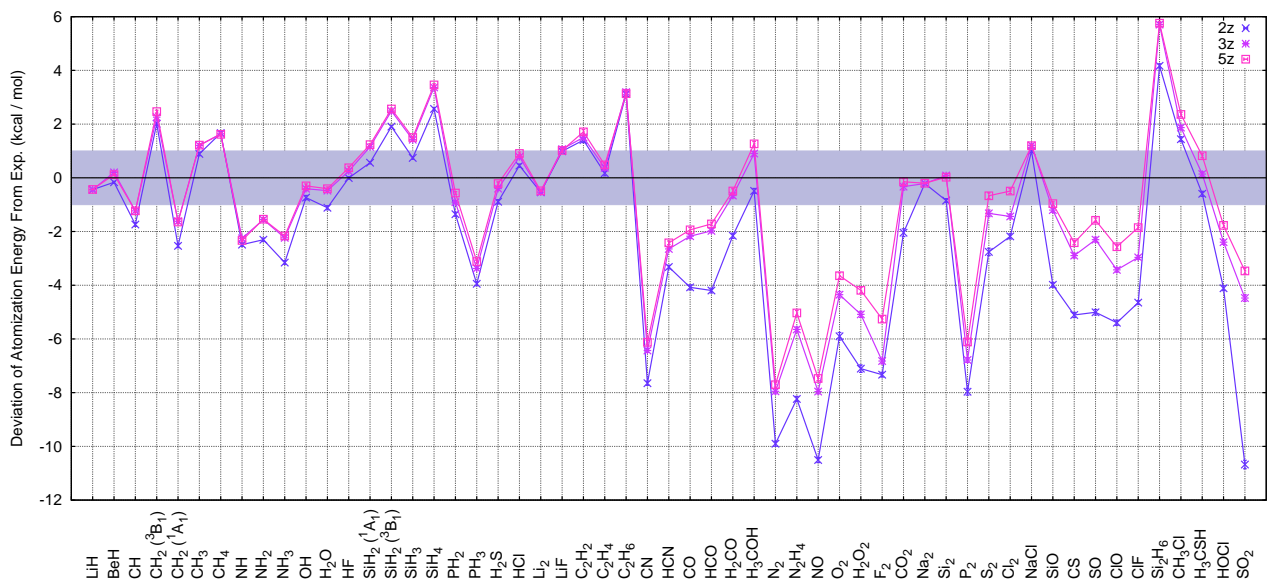}
   \caption{Deviation of the DMC atomization energies from experiment for a single determinant SJ trial wavefunction composed of VMC optimized orbitals.
     The MAD from experiment for the $2z$, $3z$, and $5z$ bases are 3.1 kcal/mol, 2.3 kcal/mol, and 2.1 kcal/mol, respectively.}
   \label{fig:jo}
 \end{center}
\end{figure*}

Although orbitals from a quantum chemistry calculation are a reasonable starting point for a QMC calculation, they are certainly not optimal due to the presence of a Jastrow factor in the QMC wavefunction.
Consequently, more accurate results are obtained by optimizing the orbitals in VMC.
The deviation of the DMC atomization energies from experiment for a single determinant SJ trial wavefunction composed of VMC optimized orbitals is shown in Figure \ref{fig:jo}.
Again, the results for $2z$, $3z$, and $5z$ basis sets demonstrate the convergence of the atomization energies with respect to basis size.
The MAD from experiment for the $2z$, $3z$, and $5z$ bases are 3.1 kcal/mol, 2.3 kcal/mol, and 2.1 kcal/mol, respectively.

\begin{figure*}[htp]  
 \begin{center}
   \includegraphics[scale=0.45]{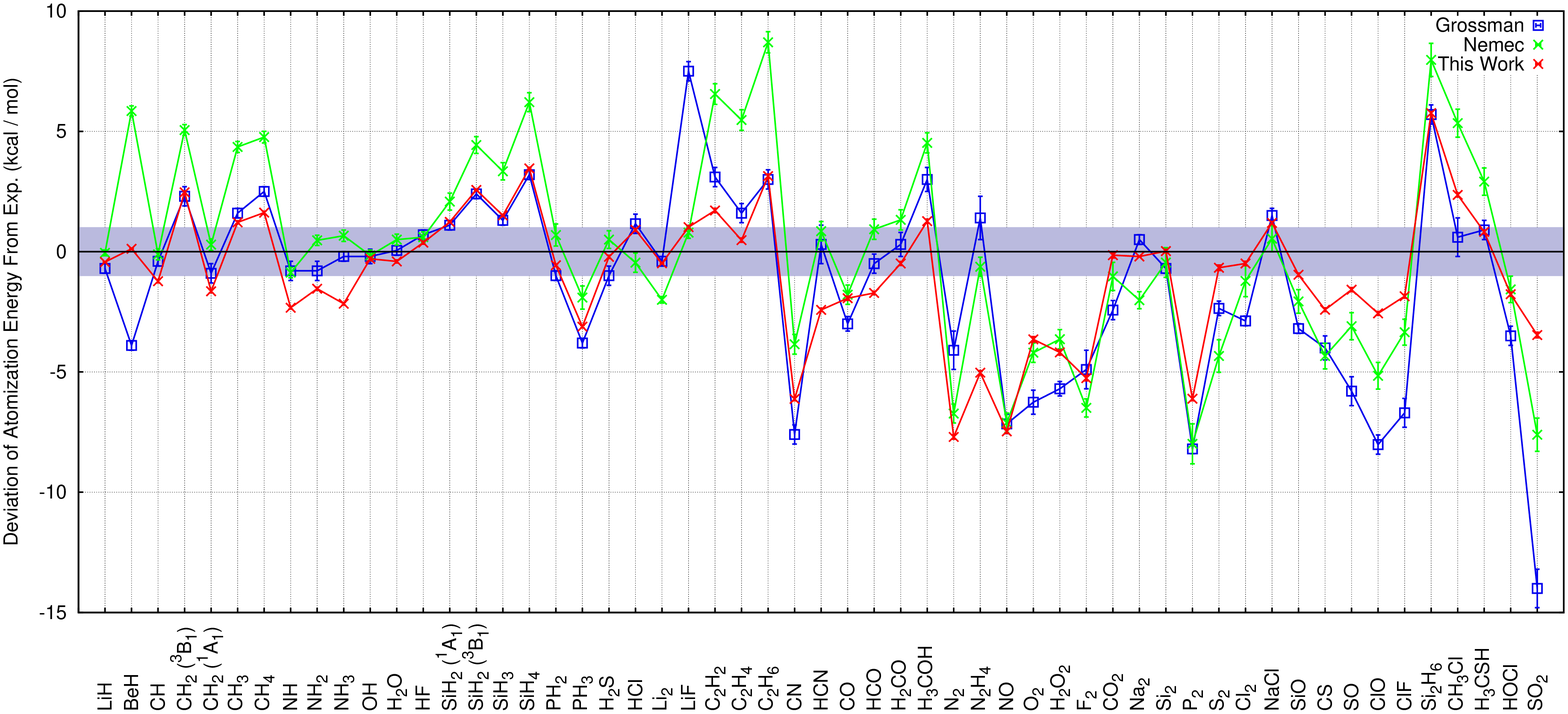}
   \caption{Comparison of the deviation of the DMC atomization energies from experiment for a single determinant SJ trial wavefunction.
     The results from this work are for a $5z$ basis and VMC optimized orbitals.
     The MAD from experiment for this work is 2.1 kcal/mol.
     The results of Nemec et al. and Grossman \cite{Grossman2002,Nemec2010} were obtained with HF orbitals and MCSCF natural orbitals, respectively.
     The MAD from experiment for Nemec et al. and Grossman are 3.1 and 2.9 kcal/mol, respectively.}
   \label{fig:vs_previous}
 \end{center}
\end{figure*}

As seen in Figure \ref{fig:vs_previous}, the orbital optimized results are noticeably better than previous QMC studies \cite{Grossman2002,Nemec2010} which produce a MAD from experiment of about 3.0 kcal/mol.
The gains in MAD from orbital optimization are 1.4 kcal/mol, 0.9 kcal/mol, and 0.9 kcal/mol for the three bases, respectively.
Although, the largest gain is for the $2z$ basis, it is evident that the benefits of orbital optimization remain for even the largest basis set.
It is worth pointing out that using optimized orbitals and a $2z$ basis produces results of similar quality to HF orbitals with a $5z$ basis.

\begin{figure*}[htp]  
 \begin{center}
   \includegraphics[scale=0.45]{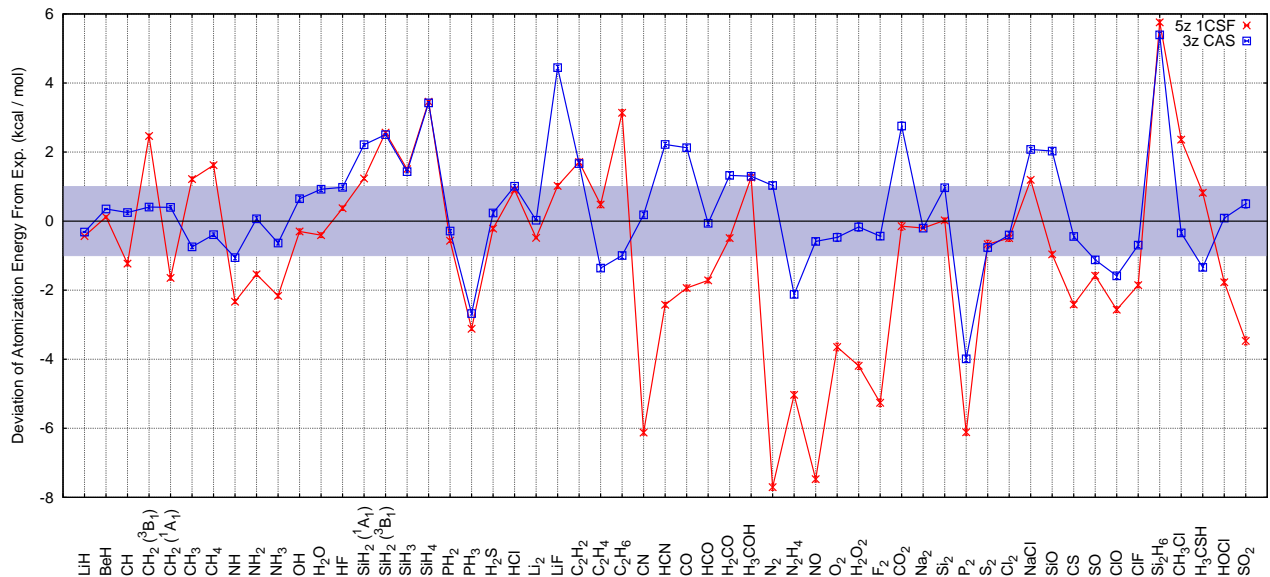}
   \caption{Deviation of the DMC atomization energies from experiment for a single determinant SJ trial wavefunction composed of VMC optimized orbitals and a CAS SJ trial wavefunction.
     The MAD from experiment for the single determinant SJ trial wavefunction is 2.1 kcal/mol.
     The MAD from experiment for the CAS SJ trial wavefunction is 1.2 kcal/mol.}
   \label{fig:1csf_to_cas}
 \end{center}
\end{figure*}

Although orbital optimization provides significant improvements to the atomization energy, the results are still a long way off from chemical accuracy.
To approach chemical accuracy, it is necessary to move beyond a single determinant SJ trial wavefunction because orbital optimization alone
does not provide sufficient flexibility in the nodal surface of the trial wavefunction.
Since the MAD of atomization energies from experiment for the $3z$ basis is only 0.2 kcal/mol higher than that of the $5z$ basis, 
and the cost of performing orbital optimization scales quadratically with the number of basis functions,
the $3z$ basis used here represents a compromise between accuracy and computational efficiency.
The deviation of the DMC atomization energies from experiment for the $s$ and $p$ valence CAS SJ trial wavefunctions is shown in Figure \ref{fig:1csf_to_cas}.
The $5z$ single determinant results are included to demonstrate the benefit of using a CAS SJ trial wavefunction.
This modest basis and CSF expansion results in a MAD from experiment of 1.2 kcal/mol, a significant step forward for QMC.

As seen with the single determinant SJ results, both increasing the basis size and optimizing the orbitals have the effect of increasing the atomization energies for every system,
since the energy gain is larger for the molecule than its constituent atoms. Since the small basis, single determinant SJ DMC results in most systems
are underbound, this on average reduces the MAD of the atomization energies. On the other hand, going from single determinant to CAS trial wavefunctions
increases the atomization energies for some systems and decreases it for others, but on average in the correct direction to reduce the MAD.
For example, the atomization energies of CH and CH$_2$($^1$A$_1$) are increased and that of CH$_2$($^3$B$_1$) reduced, but all of these changes result
in better agreement with experiment.
However, using the CAS trial wavefunctions certainly does not always improve agreement with experiment, e.g. LiF and CO$_2$.

\begin{figure}[htp] \begin{center}
   \includegraphics[scale=0.45]{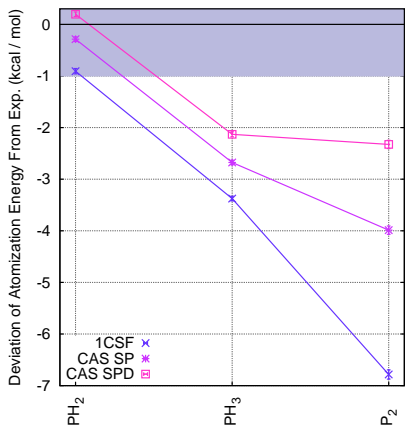}
   \caption{Deviation of the DMC atomization energies from experiment for a single determinant SJ trial wavefunction composed of VMC optimized orbitals, a CAS SJ trial wavefunction with an $s$ and $p$ active space,
     and a CAS SJ trial wavefunction with an $s$, $p$, and $d$ active space.
     The MAD from experiment for the phosphorous containing systems of the G2 set with these trial wavefunctions is  3.7, 2.3, and 1.6, respectively.}
   \label{fig:p_systems_vs_cas}
 \end{center}
\end{figure}

QMC can do yet better.
Using a larger active space will certainly help, as the largest impediment for QMC is the fixed-node error.
The choice of the modest $s$ and $p$ valence CAS allows for the possibility of scaling up to larger systems.
However, for some systems an $s$ and $p$ valence CAS may not be sufficient to properly describe the nodal structure.
To explore this, further study is performed on the phosphorous containing systems of the G2 set: PH$_2$, PH$_3$, P$_2$.
Each of these systems is underbound for the $s$ and $p$ valence CAS.
As shown in Figure \ref{fig:p_systems_vs_cas}, using $s$, $p$, and $d$ valence CAS improves agreement between DMC atomization energies and experiment.
The MAD from experiment for these three systems is 3.7, 2.3, and 1.6, for single determinant, $s$ and $p$ valence CAS, and $s$, $p$, and $d$ valence CAS, respectively.

Although using a larger active space for the phosphorus systems is beneficial, a large active space becomes impractical as system size increases.
Even though the number of CSFs included in a QMC calculation via the dual criterion described in Section \ref{sec:setup}
is much smaller than the total number of CSFs for a given active space, it is impractical to even perform the initial MCSCF calculation for large systems.
Some options for alleviating this problem are obtaining the initial trial wavefunction from less expensive configuration interaction (CI) rather than MCSCF calculations,
or, from restricted active space rather than complete active space calculations.

It is likely that some of the deviations of our results from experiment are due to using pseudopotentials.
These deviations could be evaluated by performing a similar study with the all-electron couloumbic potential.
However, there are some advantages to using pseudopotentials too.  First, all-electron calculations for molecules containing second and higher row atoms are expensive.
Second, it is possible that the fixed-node error for a given active space is larger for all-electron calculations.
Finally, the use of pseudopotentials provides a simple way of including the scalar relativistic corrections.

Additionally, some of the deviations of our results from experiment are likely due to errors in the experimental atomization energies or zero point energies.
In particular, as seen in Figure \ref{fig:1csf_to_cas}, systems containing both Si and H systematically overbind.
Additionally, very accurate all-electron frozen-core coupled cluster calculations which produce sub-1 kcal/mol MAD from experiment for the G2 set \cite{Feller1999} also systematically overbind these systems.
In particular, Feller et al. overbinds SiH$_2$($^1$A$_1$), SiH$_2$($^3$B$_1$), SiH$_3$, SiH$_4$, Si$_2$H$_6$ by 1.3 kcal/mol, 1.1 kcal/mol, 0.2 kcal/mol, 1.6 kcal/mol, 3.5 kcal/mol, respectively.

\section{Conclusion}
\label{sec:conclusion}

A QMC study of the atomization energies for the G2 set of molecules was presented.
Basis size dependence of DMC atomization energies was studied with a single determinant SJ trial wavefunction formed from HF orbitals.
With the largest basis set, the mean absolute deviation from experimental atomization energies for the G2 set was found to be 3.0 kcal/mol, in agreement with previous QMC studies.

It was determined that optimizing the orbitals within VMC improved the agreement between DMC and experiment, reducing the mean absolute deviation to 2.1 kcal/mol.
In fact, using optimized orbitals and a $2z$ basis produced results of similar quality to HF orbitals with a $5z$ basis.

Finally, DMC results for a CAS SJ trial wavefunction were near chemical accuracy with MAD from experimental atomization energies of 1.2 kcal/mol.
Although a MAD of 1.2 kcal/mol is a significant step forward for QMC, 
this result is still lackluster compared to all-electron frozen-core coupled cluster calculations which produce sub-1 kcal/mol results for the G2 set \cite{Feller1999, Feller2008}.
Several directions for improving upon the current results are larger active spaces, backflow transformations \cite{Rios2006}, yet more accurate pseudopotentials, or all-electron calculations.

\section{Acknowledgments}
We thank Claudia Filippi for very valuable discussions pertaining to the T-moves approach for treating nonlocal pseudopotentials,
and, for sending us the revised hydrogen pseudopotential.
This work was supported in part by NSF grant DMR-0908653.
Computations were performed in part at the Computational Center for Nanotechnology Innovation at Rensselaer Polytechnic Institute.
\clearpage
\bibliographystyle{apsrev4-1}
%

\end{document}